# Nanometer Scale Electronic Reconstruction at the Interface between LaVO$_3$ and LaVO$_4$

L. Fitting Kourkoutis,[1] Y. Hotta,[2] T. Susaki,[2] H. Y. Hwang,[2,3] and D. A. Muller[1]

[1]School of Applied and Engineering Physics, Cornell University, Ithaca, NY 14853, USA

[2]Department of Advanced Materials Science, University of Tokyo, Kashiwa, Chiba 277-8561, Japan

[3]Japan Science and Technology Agency, Kawaguchi, 332-0012, Japan

Electrons at interfaces, driven to minimize their free energy, are distributed differently than in bulk. This can be dramatic at interfaces involving heterovalent compounds. Here we profile an abrupt interface between V $3d^2$ LaVO$_3$ and V $3d^0$ LaVO$_4$ using electron energy loss spectroscopy. Although no bulk phase of LaVO$_x$ with a V $3d^1$ configuration exists, we find a nanometer-wide region of V $3d^1$ at the LaVO$_3$/LaVO$_4$ interface, rather than a mixture of V $3d^0$ and V $3d^2$. The two-dimensional sheet of $3d^1$ electrons is a prototypical electronic reconstruction at an interface between competing ground states.

Recent advances in oxide thin film growth and probe techniques have revealed a host of important considerations for the electronic structure at interfaces. For example, carrier depletion has been found to limit the superconducting critical current at cuprate grain boundaries [1]. Reduced surface magnetism limits the magnetoresistance of manganite tunnel junctions [2]. In these examples, atomic scale doping of the interface was found to significantly improve device performance. The role of this doping is to compensate for deviations of the surface electronic structure from the bulk, which can often occur in response to the electrostatic boundary conditions at the interface [3]. These deviations can be large in multivalent transition metal oxides, for which significant charge shifts are energetically accessible.

In addition to being a mechanism for degrading bulklike properties in device geometries, electronic interface reconstructions also present a unique synthetic opportunity to create new artificial states between two materials. Recent examples include interface ferromagnetism induced between two paramagnets [4] and conducting interfaces induced between two insulators [5, 6]. These phenomena have stimulated theoretical efforts to examine interface phases created by charge transfer in artificial structures [7, 8]. The examples of interface phases given above utilize valence states that are accessible in bulk form. Here we demonstrate experimentally that interface reconstructions can also be used to induce and stabilize nonbulklike valence states via a similar mechanism.

The multivalent $LaVO_x$ shows a variety of electronic and magnetic properties as the vanadium valence changes from 3+ to 5+ [9-12]. $LaV^{3+}O_3$, with a $3d^2$ shell configuration, is a Mott insulator, whereas $LaV^{5+}O_4$ is a band insulator with a $3d^0$ configuration. Intriguingly, bulk LaVO$_x$ lacks a structural phase corresponding to V$^{4+}$ [13]. However, at the interface between LaV$^{+3}$O$_3$ and LaV$^{+5}$O$_5$, continuity of the charge distribution implies that the vanadium valence should transition through 4+, at least in an average sense. Microscopically, the question is whether this is achieved through a mixture of the ground state *3d$^0$* and *3d$^2$* configurations, likely to be favored by a Maxwell construction from the bulk phases, or whether a local region of *3d$^1$*-configured vanadium ions can be stabilized by the interface.

In this letter, we present a study of the microscopic electronic structure of LaVO$_x$ films by spatially resolved electron energy loss spectroscopy (EELS) in a scanning transmission electron microscope (STEM). Vanadium L edge fingerprints for V$^{3+}$, V$^{4+}$ and V$^{5+}$ were established from bulk reference samples and used for the characterization of the vanadate films and the interface between LaV$^{3+}$O$_3$ and LaV$^{5+}$O$_4$. We find that LaVO$_3$ grows epitaxially on the SrTiO$_3$ substrate, resulting in atomically flat surface regions, whereas LaVO$_4$ nucleates on a thin LaVO$_3$ layer and forms polycrystalline 3D islands. At the LaV$^{3+}$O$_3$/LaV$^{5+}$O$_4$ interface, our EELS measurements reveal a two-dimensional layer of *3d$^1$*-configured V$^{4+}$ has been stabilized, thus achieving an interface charge configuration that was not accessible in bulk LaVO$_x$.

LaVO$_x$ films were grown on TiO$_2$ terminated (001)-oriented SrTiO$_3$ substrates by pulsed laser deposition using single-phase LaVO$_4$ polycrystalline targets. The LaVO$_x$ films studied here were grown at $T_g$ = 800 °C and $P_{O2}$ of 10$^{-5}$ - 10$^{-4}$ Torr, conditions for which LaVO$_3$ and LaVO$_4$ are stabilized simultaneously, resulting in mixed phase growth [14].

The microstructures of the LaVO$_x$ films are studied by annular dark field (ADF)

imaging in a 200 kV FEI Tecnai F20-ST STEM with a minimum probe size of ~1.6 Å [15] and a convergence semiangle of (9 ± 1) mrad. The chemical and electronic structures are probed on an atomic scale using spatially resolved EELS, performed on both a vacuum generator (VG) HB-501 100 kV dedicated STEM (~10 mrad), which is equipped with a parallel electron energy loss spectrometer, and on the Tecnai F20-ST fitted with a monochromator and a Gatan imaging filter (GIF) 865-ER. The energy resolution, as measured from the FWHM of the zero loss peak, was 0.5 and 0.6 eV in these two respective cases.

Figure 1 shows V-$L_{2,3}$ and O-K EELS spectra of bulk $LaVO_3$, $SrVO_3$, and $LaVO_4$. The first two main peaks of the spectra are the V-$L_3$ and $L_2$ edges. At higher energies, the peaks are dominated by the O-K fine structure; however, extraction of the O-K edge is hampered by the proximity of the V-$L_{2,3}$ edge. Large changes of the O-K and the V-$L_{2,3}$ edge fine structure are observed, in which the changes of the V-$L_{2,3}$ edge peak position are dominated by the change of the vanadium valence from 3+ to 4+ to 5+ [16, 17]. With increasing vanadium valence, the V-$L_{2,3}$ peak position shifts by ~0.9 ($V^{4+}$) and ~1.7 eV ($V^{5+}$) towards higher energies. This spectral signature can be used to determine the vanadium valence in materials with predominantly ionic character, as the V-$L_{2,3}$ edge can be reasonably described by a superposition of the spectra corresponding to the different valence states [5, 18].

Figure 2 (a) shows a cross-sectional ADF-STEM image of a $LaVO_x$ film grown on $SrTiO_3$ viewed along a substrate [001] zone axis. As expected for these growth conditions, we find two regions in the film: The brighter regions show a uniform, well-crystallized, smooth film, whereas the darker regions appear to be 3D islands that nucleated during the growth. The composition was probed using spatially resolved EELS. Figure 2(b) shows V-$L_{2,3}$/O-K spectra for the smooth regions of the film and the 3D islands, respectively. For better comparison of the fine structure, the spectra were normalized to the integrated intensity under the V-$L_{2,3}$ edge. On the 3D islands, the V-$L_3$ peak shifts by $\Delta E \sim 1.7$ eV towards higher energies. This energy shift and the direct comparison of the spectra with the V fingerprints (Fig. 1) are consistent with a vanadium valence change from $V^{3+}$ in the smooth film to $V^{5+}$ in the 3D islands. Hence, under these growth conditions there is two-phase growth of uniform, smooth $LaVO_3$ and 3D $LaVO_4$ islands. Figure 2(c) shows an atomic force microscopy (AFM) image representative of the film, confirming the formation of 3D islands and regions of atomically flat surface, which correspond to $LaVO_4$ and $LaVO_3$, respectively.

The reduced contrast of the $LaVO_4$ islands compared to the $LaVO_3$ layer is due to the relative orientation of the crystals compared to the electron propagation direction. ADF-STEM images of crystals are sensitive to crystal orientation [19], because channeling of the probe on the atom column for crystals oriented on zone axis leads to contrast enhancement in the ADF-STEM image. As the crystal is tilted off axis, the channeling effect is reduced and the contrast diminishes. In Fig. 2, the $LaVO_3$ and the $SrTiO_3$ substrate are aligned on zone axis, whereas the $LaVO_4$ polycrystalline islands are mostly oriented off the axis. Figure 3(c) shows a high resolution ADF-STEM image of a region in a $LaVO_4$ island that is oriented on the zone axis. The difference between the $LaVO_4$ [Fig. 3(c)] and the $SrTiO_3$ crystal structure [Fig. 3(d)] is apparent. $SrTiO_3$ shows the typical cubic perovskite structure, whereas $LaVO_4$ is monoclinic with monazite-type structure ($P2_1/c$, a=7.047 Å, b=7.286 Å, c=6.725 Å, β=104.85°) [20], viewed in Fig. 3(c) along the LaVO$_4$ [010] direction.

Figure 3(b) and 3(d) show a bright band between the LaVO$_4$ and the SrTiO$_3$ substrate, which suggests a chemical change for zone-oriented crystals. EELS on the bright layer confirms the presence of LaVO$_3$; i.e., the 3D LaVO$_4$ islands do not nucleate directly on the SrTiO$_3$ substrate but on a thin (~2nm) LaVO$_3$ layer. This raises an interesting question of how the microscopic transition from V$^{3+}$ to V$^{5+}$ is made at this abrupt interface. To explore this, we have measured V-L$_{2,3}$ edge spectra across the interface between the thin LaVO$_3$ layer and the LaVO$_4$ island. Figure 4 shows the result of a scan across the interface. The probing path is indicated by the dotted line in the ADF image [Fig. 4(c)]. By decomposing the V-L$_{2,3}$ near edge structure into a linear combination of V$^{3+}$, V$^{4+}$ and V$^{5+}$ components [Fig. 4(b)], we can monitor the change of V valence at the interface. The results of this three-component fit are shown in Figs. 5(a) and 5(c). As expected, we find the transition from V$^{3+}$ in the thin bright layer to V$^{5+}$ in the LaVO$_4$ islands. However, at the crossover between these valence states, we find a significant contribution of V$^{4+}$ (i.e., V in a *3d$^1$* configuration), with an area accounting for close to one electron at the LaVO$_3$/LaVO$_4$ interface. The residuals, i.e., the root-mean square deviation of the experimental data from the fit, are constantly low across the interface in the three-component case. A fit of this series of spectra with only two components (V$^{3+}$ and V$^{5+}$) as shown in Fig. 5(b) results in large residuals at the interface, indicating the presence of a third component. Hence, at the LaVO$_3$/LaVO$_4$ interface the vanadium valence varies smoothly from V$^{3+}$ to V$^{5+}$, passing through a region of V$^{4+}$ at the interface.

Microscopically, the local electronic distribution is not phase separated into only the bulk-accessible $3d^0$ and $3d^2$-configured V d states but also includes a thin layer of $3d^1$ V atoms at the interface. The $3d^1$ state would not be expected from the bulk available phases, nor is it required from continuity of the wave function in an average sense, highlighting the very different electronic response at the interface.

In summary, we have shown that the nonbulklike charge state $3d^1$ V can be stabilized at the $LaVO_3$/$LaVO_4$ interface. Here an interface between two competing ground states (the $3d^2$ Mott insulator and $3d^0$ band insulator) has led to a two-dimensional sheet of intermediate valence $3d^1$ electrons. As there is a large family of closely lattice-matched transition metal oxides with different and exotic ground states, other electronic reconstructions and intermediate valence states may also be stabilized by this approach..

We thank M. Nohara for helpful discussions. L.F.K. and D.A.M. acknowledge support under the ONR EMMA MURI monitored by Colin Wood, and by the Cornell Center for Materials Research (NSF No. DMR–0520404 and No. IMR-0417392). Y.H., T.S., and H.Y.H. acknowledge support from a Grant-in-Aid for Scientific Research on Priority Areas. L.F.K. acknowledges financial support by Applied Materials. Y.H. acknowledges support from QPEC, Graduate School of Engineering, University of Tokyo.

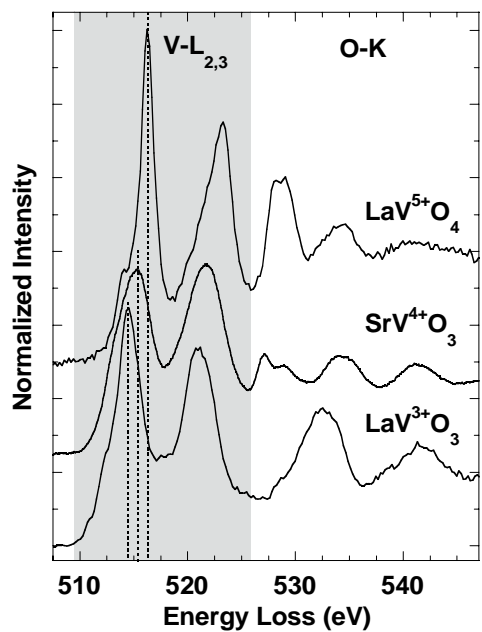

FIG. 1. V-$L_{2,3}$ and O-K electron energy loss spectra taken from bulk $LaVO_3$, $SrVO_3$ and $LaVO_4$. The dashed lines indicate the $L_3$ peak position, which shifts to higher energies as the vanadium valence increases. The shaded region of the spectra is due to V-$L_{2,3}$ excitations only, whereas at higher energies the EELS signal is composed of V and O-K edge contributions. The measurements were performed on the FEI Tecnai F20-ST equipped with a GIF 865-ER.

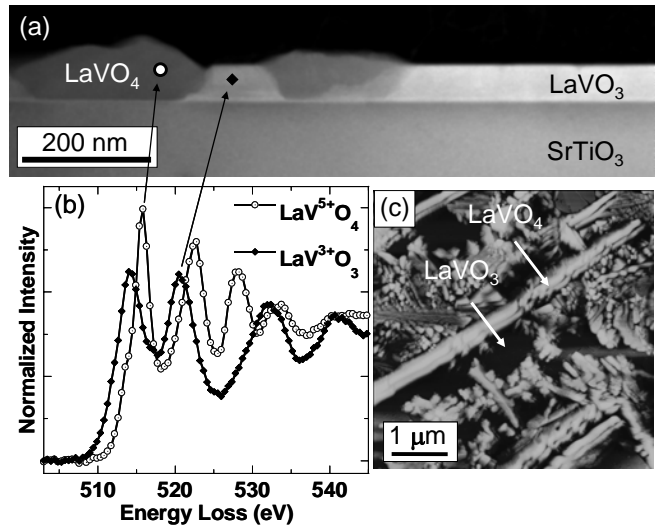

FIG. 2. (a) Cross-sectional ADF image of a LaVO$_x$ film grown at T$_g$=800°C and P$_{O2}$=10$^{-4}$ Torr showing mixed phase growth of 3D LaVO$_4$ islands and flat LaVO$_3$ on a SrTiO$_3$ substrate. (b) V-L$_{2,3}$/O-K electron energy loss spectra taken from an island and from the smooth film, respectively. (c) AFM image of the same film showing both flat regions of LaVO$_3$ and 3D islands of LaVO$_4$.

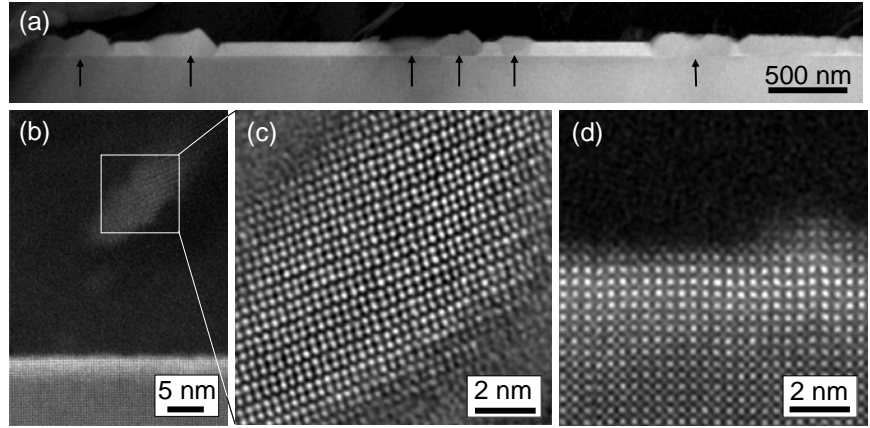

FIG. 3: (a) Low magnification ADF image of a LaVO$_3$ film grown at T$_g$=800°C and P$_{O2}$=10$^{-4}$ Torr, where LaVO$_4$ 3D islands are marked with arrows. (b) Higher magnification ADF image of a LaVO$_3$ film grown at T$_g$=800°C and P$_{O2}$=10$^{-5}$ Torr showing a bright band at the interface between the LaVO$_4$ and the SrTiO$_3$ substrate. (c) The crystal structure of the monoclinic LaVO$_4$ recorded along the [010] crystallographic direction for the area marked by the box in (b). (d) The sharp interface between the LaVO$_4$ and the bright band.

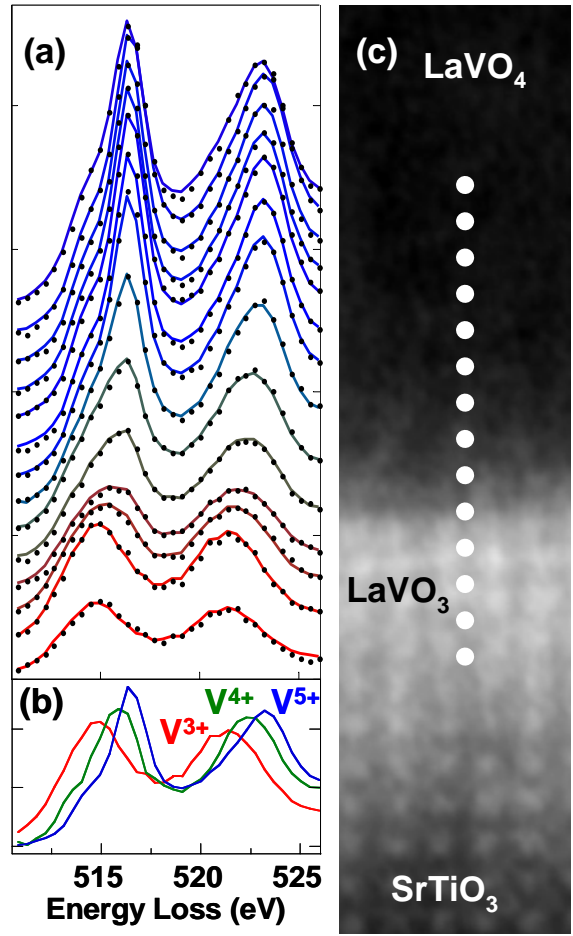

FIG. 4: (a) V-$L_{2,3}$ electron energy loss spectra recorded across the $LaVO_4/LaVO_3$ interface (dotted lines). The solid lines show the fits to the experimental spectra using reference spectra shown in (b). (c) ADF image of the interface where the electron beam position is denoted by the white dotted line. The measurements (a), (c) were performed on the VG HB-501 100 kV dedicated STEM equipped with a parallel electron energy loss spectrometer.

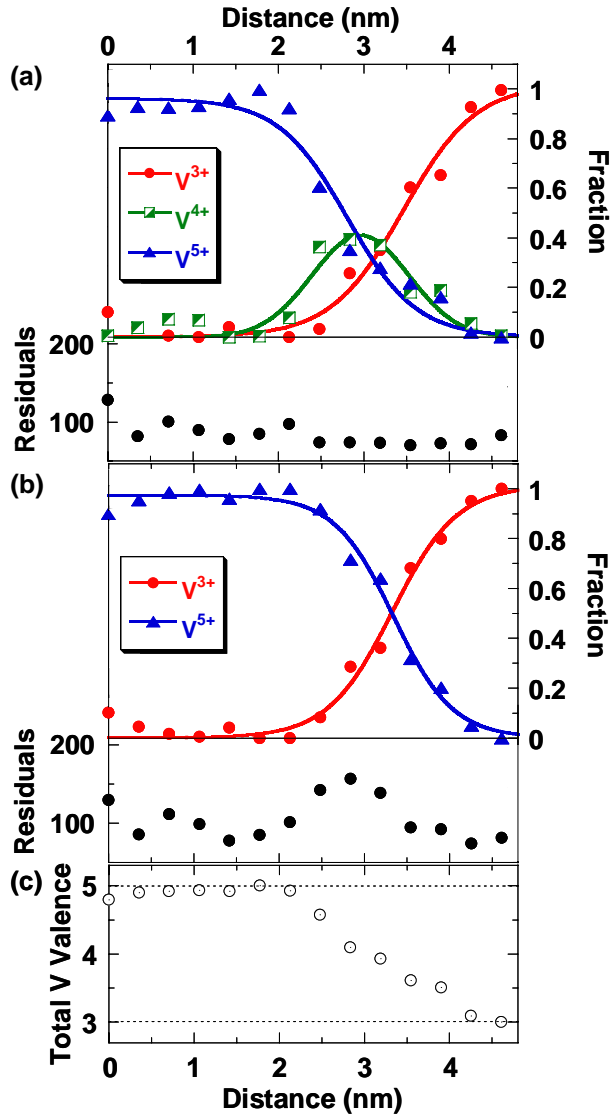

FIG. 5: (a) The fractions of $V^{3+}$, $V^{4+}$ and $V^{5+}$ across the interface obtained by a three-component fit to the experimental data shown in Fig. 4(a) using the reference spectra of Fig. 4(b). These fractions were used to color code the fits to the experimental data in Fig. 4(a). The root-mean square residuals of the fit are shown below. (b) A two-component fit of the data using $V^{3+}$ and $V^{5+}$ reference spectra results in large residuals at the interface. (c) Total vanadium valence across the $LaVO_4/LaVO_3$ interface as determined from (a).